\documentclass[conference]{IEEEtran}
\IEEEoverridecommandlockouts
\usepackage{cite}
\usepackage{amsmath,amssymb,amsfonts}
\usepackage{algorithmic}
\usepackage{graphicx}
\usepackage{textcomp}
\usepackage{xcolor}
\usepackage{float}
\usepackage{placeins}
\usepackage{caption}
\usepackage{booktabs}

\def\BibTeX{{\rm B\kern-.05em{\sc i\kern-.025em b}\kern-.08em
    T\kern-.1667em\lower.7ex\hbox{E}\kern-.125emX}}
\begin{document}

\title{Evaluating Performance Characteristic of Opportunistic 
Routing Protocols: A Case Study of the 2016 Italian 
League Match Earthquake in the Stadio Adriatico\\

}

\author{\IEEEauthorblockN{\textsuperscript{} Yihang Cao}
\IEEEauthorblockA{\textit{School of Computer Science} \\
\textit{University of Nottingham} \\
alyyc111@nottingham.ac.uk}
\and
\IEEEauthorblockN{\textsuperscript{} Milena Radenkovic }
\IEEEauthorblockA{\textit{School of Computer Science} \\
\textit{University of Nottingham}\\
milena.radenkovic@nottingham.ac.uk }
}

\maketitle

\begin{abstract}
 Delay Tolerant Networks (DTNs) can provide emergency communication support when conventional infrastructure is disrupted during disasters. This paper evaluates the performance of opportunistic routing protocols in a realistic disaster scenario based on the 2016 Central Italy earthquake, modelled as an emergency occurring during a football match at Stadio Adriatico in Pescara. We identify multiple suitable  groups of mobile and static nodes, such as audiences, a range of different emergency responders, stage sensors, and vehicles, to design and build evacuation and rescue activities in a partially connected environment. Two representative DTN routing protocols, Epidemic and Spray and Wait, are tested under identical simulation settings and compared using delivery probability, latency, overhead ratio, hop count and dropped messages. The results highlight that Spray and Wait provides a better balance between reliability and efficiency in this scenario, achieving higher delivery probability while reducing overhead and using network resources more efficiently. The study shows the usefulness of DTN simulation for analysing disaster communication performance in emergency response scenarios.
\end{abstract}

\begin{IEEEkeywords}
Delay Tolerant Networks, disaster communication, opportunistic routing, Epidemic, Spray and Wait
\end{IEEEkeywords}

\section{Introduction}
Natural disasters are frequently accompanied by disruptions to traditional 
communication networks, especially in urbanized areas where an emergency case can 
trigger extensive use of networks \cite{b1}\cite{b4}. The failure or overloading of cellular networks blocks 
the exchange of important information so that only a few witnesses and rescuers are 
able to communicate amongst themselves. Such gaps in communications delay rescue 
efforts and hinder the effective coordination of remaining resources. Delay-Tolerant 
Networking (DTN) \cite{b1}, based on the store-carry-forward paradigm, 
introduces a viable alternative by offering opportune communication sessions in 
locations where temporary end-to-end connections cannot be assured \cite{b5}.

Basically, large public events represent a particularly challenging setting for emergency communication. The disruption associated with the 2016 Central Italy Earthquake illustrates how sudden seismic events can affect football matches and generate immediate communication demand among large crowds. In a stadium environment, high population density, abrupt mobility changes and panic-driven movement together create conditions in which conventional end-to-end communication cannot be guaranteed. These characteristics make stadium emergencies a relevant and challenging scenario for evaluating DTN-based communication support. Fig.1 indicates the 2016 Central Italy earthquake affected a broad area across central Italy \cite{b2}. Fig.2 shows the scene of earthquake during Seria A football match \cite{b3}. 
\begin{center}
\includegraphics[width=0.55\columnwidth]{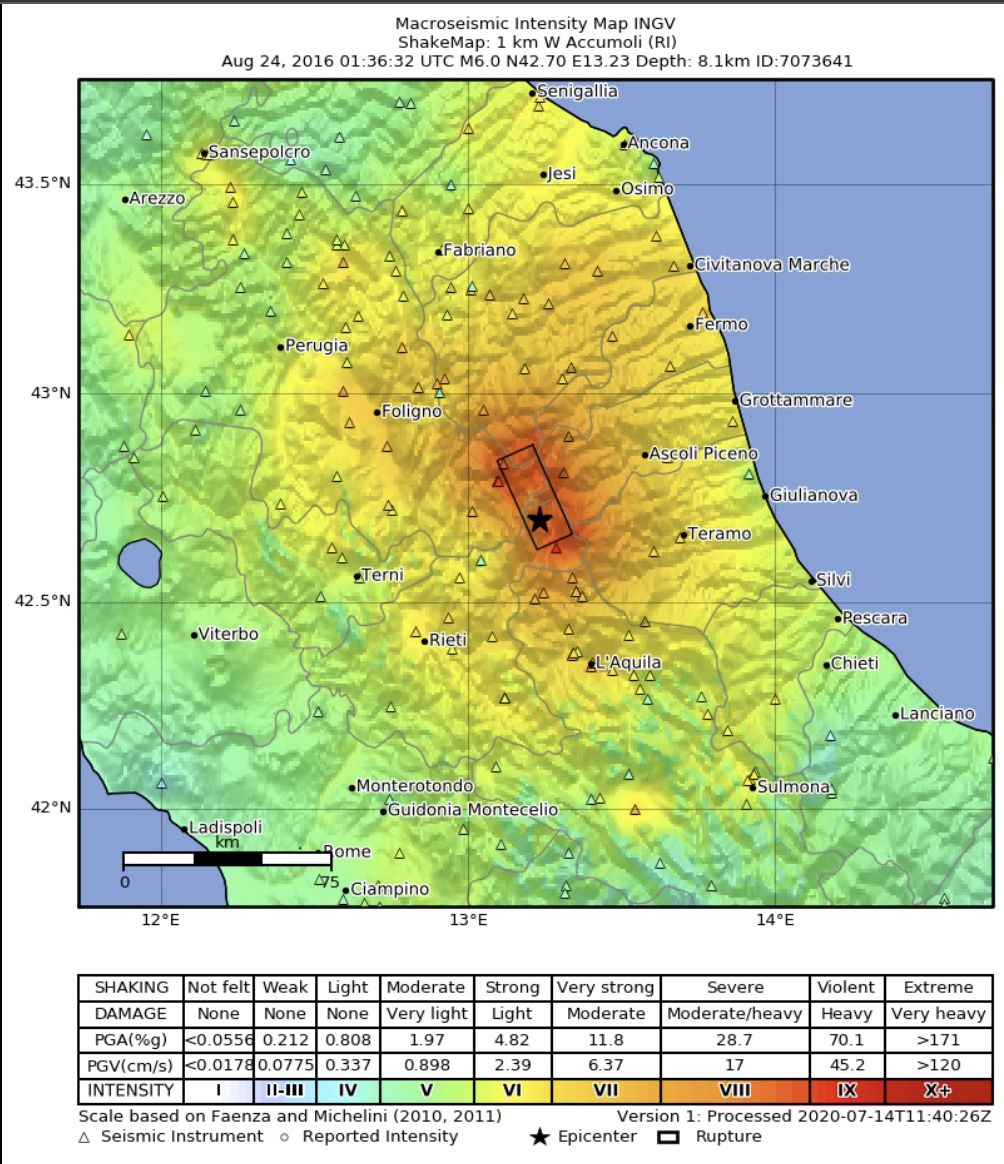}
\captionof{figure}{Macroseismic intensity map of the 2016 Central Italy earthquake.}
\label{fig1}
\end{center}

\begin{center}
\includegraphics[width=0.75\columnwidth]{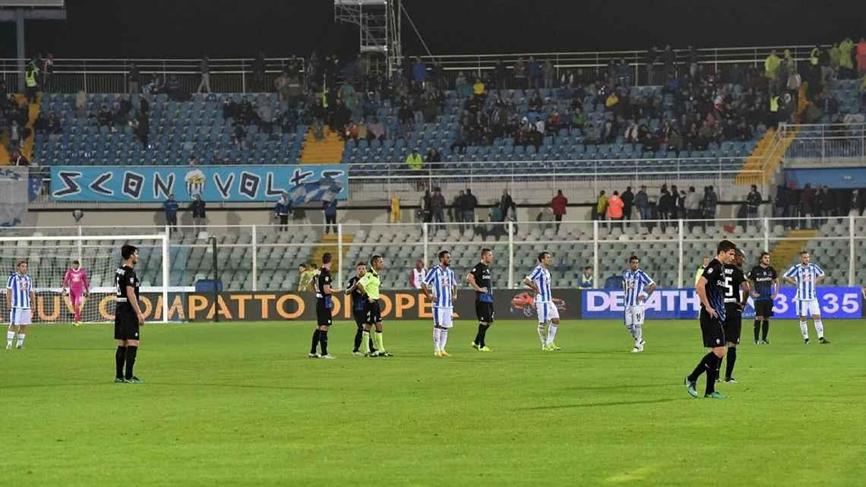}
\captionof{figure}{The scene of earthquake during Italian Serie A football match in the Stadio 
Adriatico.}
\label{fig1}
\end{center}

Prior studies have examined DTN routing in disrupted and poor infrastructure environments. However, less attention has been given to dense football-stadium earthquake scenarios with heterogeneous node roles and explicit resource constraints. In particular, the performance trade-off between flooding-based and controlled-replication routing under different buffer sizes remains insufficiently explored in this type of emergency setting. A more focused investigation of how routing strategy and storage capacity interact in a high-density disaster communication environment needed to be done.

Using the ONE simulator \cite{b9}, we model a football stadium environment involving audience members, rescue teams, ambulances, media personnel, sensors, and exits, and compare Epidemic and Spray and Wait routing using delivery probability, latency, overhead ratio, hop count, and dropped messages. This paper investigates the performance of opportunistic routing in a stadium earthquake emergency scenario, with particular emphasis on the trade-off between reliability and resource efficiency under buffer constraints.

\section{Related Work}
Delay Tolerant Networks(DTNs) have been widely studied for in situations with disconnections and with an absent or damaged infrastructure \cite{b1} \cite{b5} \cite{b6}. Foundational DTN research established the store-carry-forward paradigm as an effective solution for challenged networking conditions, and subsequent studies have identified disaster response as one of its most important application areas \cite{b4} \cite{b12}.

Epidemic routing is one of the most widely studied forwarding approaches and is valued for its robustness in sparse or unpredictable contact environments because aggressive replication increases the probability of eventual delivery \cite{b7}\cite{b14}. However, the same flooding mechanism also creates substantial overhead, rapid buffer occupancy, and severe congestion, especially in crowded or mobility-intensive disaster settings \cite{b13}\cite{b14}. To address these limitations, controlled-replication approaches such as Spray and Wait were proposed \cite{b8}. By limiting the number of message copies, Spray and Wait seeks to maintain delivery capability while reducing storage pressure and transmission cost, making it attractive in resource-constrained emergency environments. More recent studies have explored Protocols such as CafRep \cite{b15} and CafRepCache \cite{b10} attempt to improve DTN efficiency by exploiting encounter history, regional congestion information, or predicted content demand.
\section{Methodology}
In this section, it presents the modelling of the earthquake evacuation scenario for Stadio Adriatico, the simulation setup in the ONE simulator \cite{b9}, and how the comparative performance of two DTN routing protocols: Epidemic and Spray and Wait was evaluated.
\subsection{Earthquake In Football Stadium Emergency Scenario Design}
We model an earthquake emergency during a Serie A football match. The scenario design is inspired by the real event which happened in the 2016 Central Italy Earthquake. In addition, it represents a dense public venue in which conventional communication infrastructure is assumed to be unreliable or partially unavailable.Within this setting, audience nodes generate distress messages that are intended for rescue teams operating inside the stadium. The scenario is designed to capture the combined effects of crowd density, heterogeneous mobility, and intermittent connectivity on opportunistic message delivery. To evaluate the performance of selected protocols Epidemic and Spray and Wait in realistic earthquake scenarios, we constructed the ONE simulator environment based on the Stadio Adriatico. In the ONE simulator, the audiences, Media personnel,rescue teams and ambulances were designed according to the ShortestPathMapBasedMovement model. These mobile nodes were not assigned a predetermined exit route. This assumption reflects the uncertainty and disruption that typically characterise earthquake emergencies, where orderly evacuation information may be unavailable or delayed. The ONE simulator was selected because it is specifically designed for evaluating delay-tolerant and opportunistic networking scenarios. It provides built-in support for map-based mobility, heterogeneous node configuration, message-level routing simulation, and standard DTN performance metrics such as delivery probability, latency, and overhead ratio. These capabilities make it well suited to modelling a disrupted stadium environment and to comparing the behaviour of Epidemic and Spray-and-Wait under consistent experimental conditions. Fig.3 displays the simulated scenario GUI when we run the ONE simulator.
\begin{center}
\includegraphics[width=0.8\columnwidth]{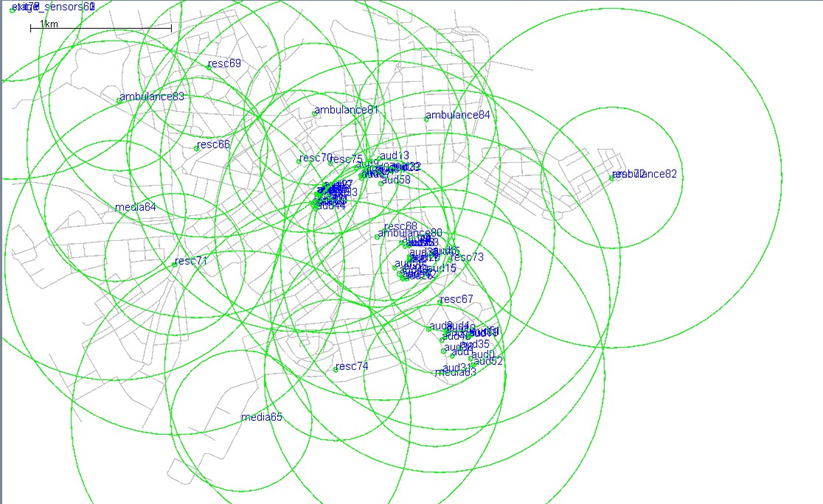}
\captionof{figure}{Earthquake emergency scenario model in the ONE simulator.}
\label{fig1}
\end{center}

\subsection{Node Groups and Mobility Assumptions}
Six node groups are defined to represent the functional diversity of a stadium emergency environment: audiences, stage sensors, media personnel, rescue teams, exits, and ambulance units. And a total number of 85 mobile nodes were configured. Audience nodes act as the main message sources, reflecting injured or distressed spectators seeking assistance. Rescue teams are modelled as the primary destinations for these messages and move through the venue to coordinate response activities. Ambulance nodes represent highly mobile medical responders, while media personnel serve as fast-moving opportunistic carriers. Stage sensors and exit nodes are treated as stationary or semi-static infrastructure elements that can support local relaying and monitoring. This heterogeneous node design enables the simulation to capture different mobility patterns and forwarding opportunities within the same disrupted environment. The specific details about the number of node groups design are shown in Fig.4.
\begin{center}
\includegraphics[width=0.65\columnwidth]{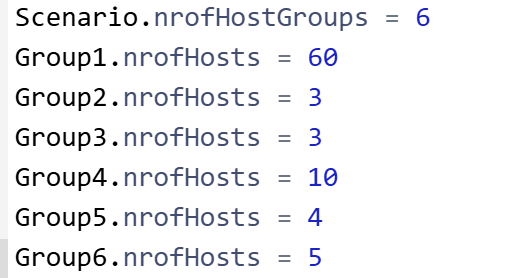}
\captionof{figure}{Number of node groups design.}
\label{fig1}
\end{center}

 The mobility is configured to reflect role-specific behaviour. Under evacuation pressure, audience nodes move slowly to simulate dense crowd flow, with the speed between 0.4-1.0 m/s, using ShortestPathMapBasedMovement model. Due to the crowd congestion, the nodes may pause for some time before moving again. Fig.5 illustrates the design of audience group nodes.
 \begin{center}
\includegraphics[width=0.8\columnwidth]{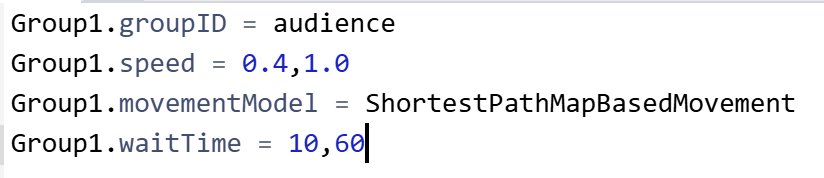}
\captionof{figure}{The design of audience group nodes.}
\label{fig1}
\end{center}

whereas, rescue teams and ambulances are assigned higher speeds to represent active emergency response. The mobility model employs ShortestPathMapBasedMovement model, ensuring their higher mobility and operational priority. And their movement is still constrained by the physical layout of the stadium and surrounding road network. Rescue team nodes are configured with movement speed(2.0-5.0 m/s). The ambulance units are configured with higher movement speed between 3.0 m/s and 12.0 m/s. Fig.6 shows the design of rescue teams and ambulance units group nodes.

\begin{center}
\includegraphics[width=0.8\columnwidth]{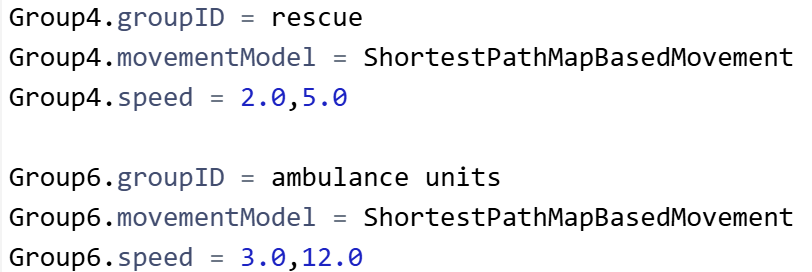}
\captionof{figure}{The design of rescue teams and ambulance units group nodes.}
\label{fig1}
\end{center}

Media nodes occupy an intermediate position, as they may move rapidly toward incident locations but do not share the same operational priority as rescue personnel. Fig.7 shows the design of media group nodes.

\begin{center}
\includegraphics[width=0.8\columnwidth]{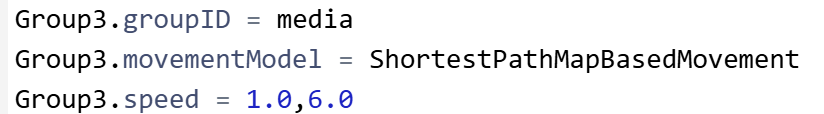}
\captionof{figure}{The design of media group nodes.}
\label{fig1}
\end{center}

In contrast, stage sensors and exits remain stationary, thereby providing fixed points within the communication topology. As a fixed high-performance node, it can act as a highly stable relay node in the network, helping to increase the transmission success rate between the rescue team and the audience. The exit nodes represent the main evacuation points within the stadium and they assist the rescue team in monitoring the evacuation situation. Fig.8 displays the design of stage sensors and exits group nodes.

\begin{center}
\includegraphics[width=0.8\columnwidth]{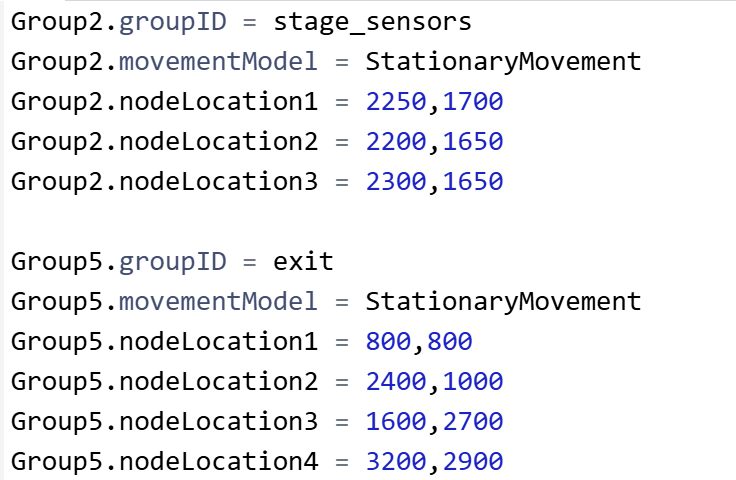}
\captionof{figure}{The design of stage sensors and exits group nodes.}
\label{fig1}
\end{center}

\subsection{Communication and Traffic Model}
For the emergency scenario communication design, the message generation follows a period process. The messages
were generated every 30 to 60 seconds and the size of them ranged from 100kB to 300kB. Audience nodes act as the primary message senders, While the rescue teams serve as the crucial receivers. This reflects a normal and important communication pattern. Each message is assigned a time-to-live (TTL) of 3 hours to ensure message validity throughout the crucial rescue window. Otherwise, it represents the operational relevance of urgent information during the emergency response window.

Buffer size is the main experimental variable and is varied across 5MB, 10MB, 15MB, 20MB. This decision is motivated by the central role of storage pressure in opportunistic networking, especially for protocols that rely on replication. By changing the buffer capacity while keeping the other conditions unchanged, this simulation isolates how storage availability affects routing performance like reliability, latency, and overhead and so on. To simulate the wireless communication, we use three interface types of SimpleBroadcastInterface. All mobile and fixed nodes are modelled with the Bluetooth interface. Additionally, stage sensor, media, rescue team, exit and ambulance units group are applied the higher-speed, bigger-range interface like wifiInterface and highspeedInterface. Detailed 
simulation parameters are presented in Table 1.
\begin{table}[htbp]
\centering
\caption{Simulation parameters}
\label{tab:params}
\begin{tabular}{lll}
\toprule
Parameter & Value \\
\midrule
Simulation time & 12 h  \\
Buffer size & 5--20 MB \\
Message size & 100--300 KB  \\
Message TTL & 3 h \\
Total Nodes & 85 \\
Node Groups & 6 \\
Interface & Bluetooth(250kB/s, 15m)\\&Wifespeed(10MB/s, 500m)\\& highspeed(20MB/s, 1200m)\\
Mobility Model &ShortestPathMapBasedMovement\\&StationaryMovement \\ 
Message Creation Interval & 30-60s \\
\bottomrule
\end{tabular}
\end{table}

\subsection{Routing Protocols and Experimental Design}
Two opportunistic routing protocols are evaluated: Epidemic and Spray and Wait \cite{b7}\cite{b8}\cite{b14}. These protocols were chosen because they represent two completely different forwarding concepts in DTN research. Epidemic routing maximises delivery opportunities through unrestricted replication, whereas Spray and Wait constrains the number of message copies and therefore emphasises resource efficiency. 

To ensure a fair comparison, both protocols were tested under the same scenario configuration, node mobility assumptions, traffic generation settings, and communication parameters. The only intentionally included experimental variables were the routing mechanism itself and the controlled adjustment of the buffer size. This design allows the study to assess how flooding-based and controlled-replication routing respond to storage constraints in a dense stadium emergency environment. For the Epidemic configuration, all groups keep using the same protocol. Fig.9 shows that.
 \begin{center}
\includegraphics[width=0.8\columnwidth]{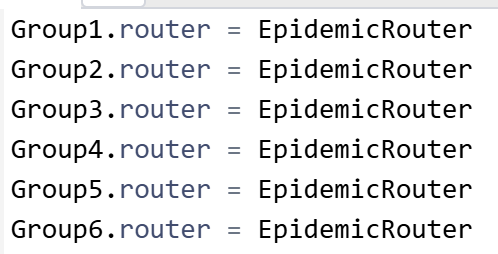}
\captionof{figure}{The design of group router protocols using Epidemic.}
\label{fig1}
\end{center}
For the Spray and Wait configuration, all groups also keep remaining the same protocol. Fig.10 illustrates that.
 \begin{center}
\includegraphics[width=0.8\columnwidth]{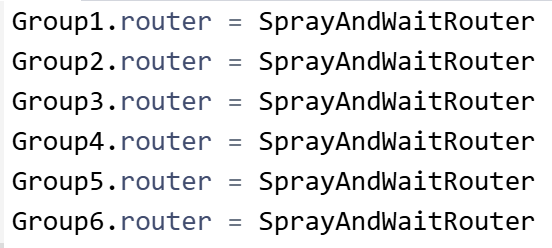}
\captionof{figure}{The design of group router protocols using Spray and Wait.}
\label{fig1}
\end{center}

\subsection{Simulation Process}
In each experimental run, audience nodes acted as the primary message source, while the rescue team served as the intended recipients in the event of a stadium emergency. After each run, simulator reports were collected and used to compare routing performance, including delivery probability, latency, overhead ratio, hop count, and the number of dropped messages. This experimental procedure enabled the separation of how storage constraints affect the relative behavior of flood-based routing and controlled replication routing in a football stadium environment affected by intense earthquakes.

\section{The evaluation of the performance in the 
emergency scenario 
 }
\subsection{Evaluation Metrics}
The main performance metrics are delivery probability, latency average, overhead ratio, hopcount average and dropped message. The comparative analysis focused on these key metrics and these metrics were selected because they provide a balanced view of both communication effectiveness and resource efficiency in a disaster scenario. 

\subsection{Results and Analysis}
Fig.11 illustrates the delivery probabilities of Epidemic and Spray and Wait routing under different buffer sizes. For Epidemic routing, the delivery probability increases with increasing buffer capacity, which is expected, as a larger buffer allows more message replicas to survive long enough to reach their destination. When the buffer size is limited to 5MB, significant buffer overflows lead to substantial message loss, resulting in a relatively low delivery probability of approximately 0.2029. Even with the buffer increased to 20MB, the delivery probability of epidemic routing only reaches approximately 0.3790, indicating that unrestricted flooding still inefficiently utilizes available storage space. In contrast, Spray and Wait maintains a higher delivery probability across all buffer settings, suggesting that controlled replication is more suitable for storage-constrained, high-density simulated stadium emergencies.
\begin{center}
\includegraphics[width=0.8\columnwidth]{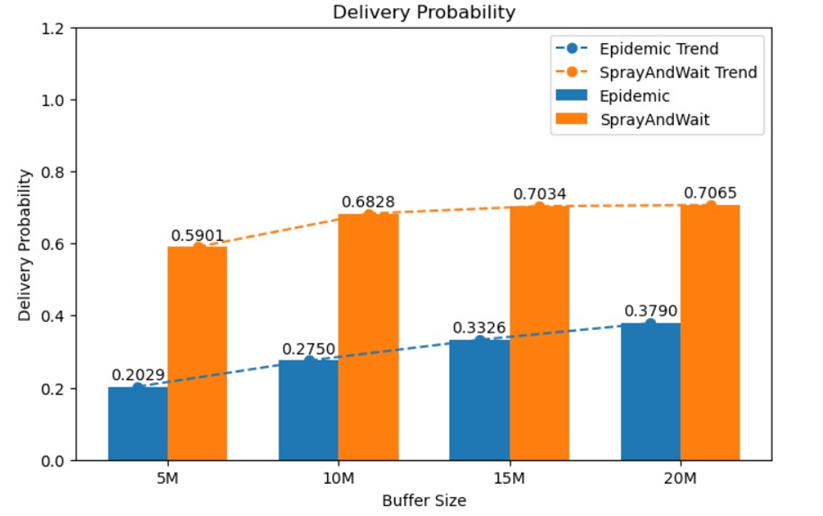}
\captionof{figure}{Delivery probability in simulated football match scenarios.}
\label{fig1}
\end{center}

Fig.12 compares the average latency of the two routing protocols. Spray and Wait shows consistently lower delay across all buffer size settings, indicating that limiting the number of message copies helps preserve shorter and less congested forwarding paths. By contrast, Epidemic routing exhibits higher latency because excessive replication increases contention for storage and forwarding opportunities. Although Epidemic latency decreases as the buffer size increases, this improvement mainly reflects the reduced severity of buffer overflow rather than fundamentally efficient routing behaviour. Overall, the latency results suggest that controlled replication is more effective than flooding when timely message delivery is required in an emergency environment.
\begin{center}
\includegraphics[width=0.8\columnwidth]{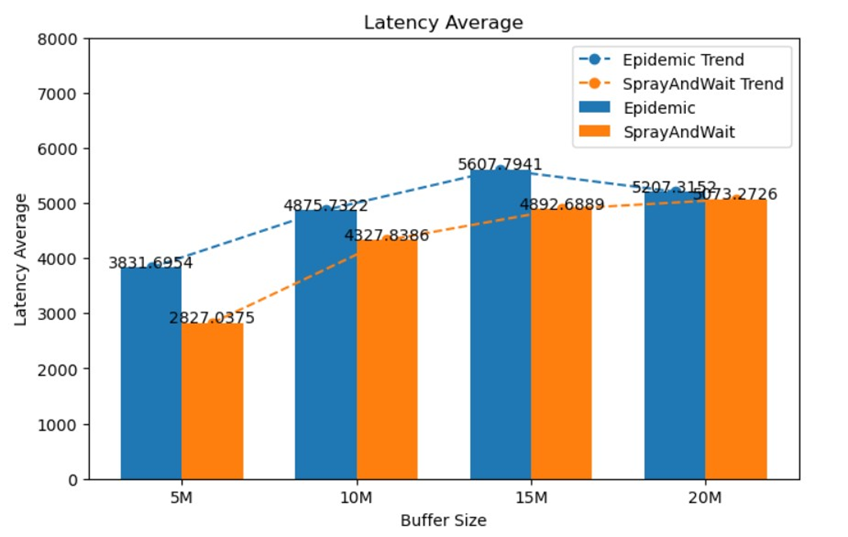}
\captionof{figure}{Latency average in simulated football match scenarios.}
\label{fig1}
\end{center}

Fig.13 shows the overhead ratio produced by both routing protocols. Epidemic routing generates dramatically higher overhead than Spray and Wait under every buffer size setting, with values ranging from approximately 2303 to 4253. This reflects the large number of redundant message copies created by uncontrolled flooding. Even when the buffer size is increased, Epidemic remains highly inefficient in terms of transmission cost. In contrast, Spray and Wait maintains a very low and stable overhead ratio, approximately between 9 and 11, because it restricts the number of replicas circulating in the network. These findings indicate that controlled replication greatly reduces communication cost and is therefore more scalable for emergency scenarios with limited storage and unstable connectivity.
\begin{center}
\includegraphics[width=0.8\columnwidth]{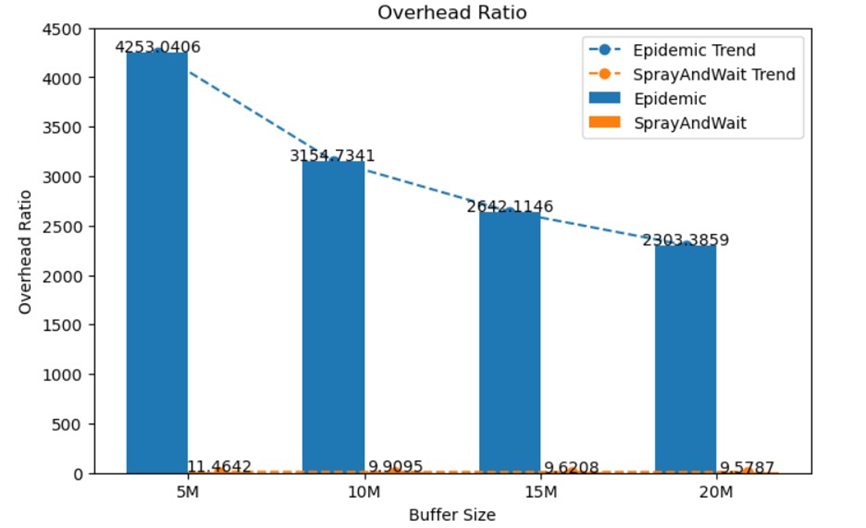}
\captionof{figure}{Overhead Ratio in simulated football match scenarios.}
\label{fig1}
\end{center}

Fig.14 presents the average hop count for delivered messages. Epidemic routing produces substantially longer paths, typically around 6 to 7 hops, because the flooding mechanism spreads messages through many redundant intermediate nodes before successful delivery. Although larger buffers reduce some duplicate loss, the routing paths remain comparatively inefficient. Spray and Wait, by contrast, maintains a stable hop count of approximately 2.7, indicating more direct and controlled forwarding. This shorter path structure helps explain its lower latency and reduced overhead. The hop count results therefore reinforce the conclusion that controlled replication provides more efficient message dissemination in the stadium earthquake scenario.
\begin{center}
\includegraphics[width=0.8\columnwidth]{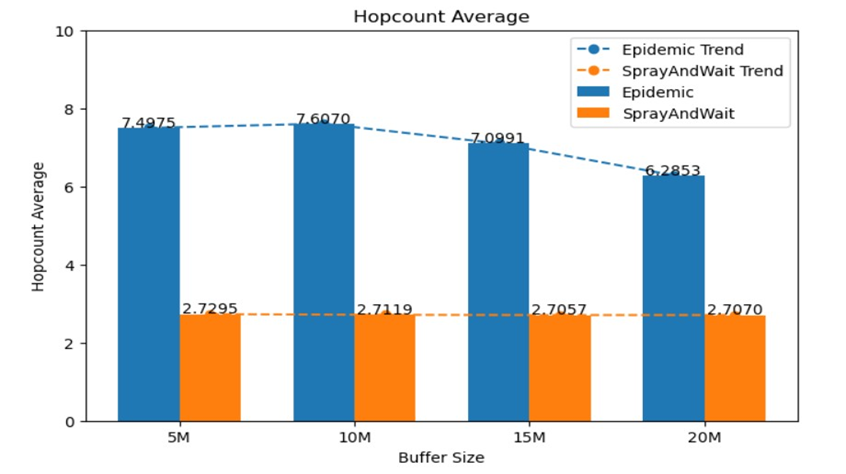}
\captionof{figure}{Hopcount Average in simulated football match scenarios.}
\label{fig1}
\end{center}
In Fig.15, Epidemic routing suffers from extremely high message drop rates, exceeding 800,000 drops under all buffer configurations. This is a direct consequence of its uncontrolled flooding strategy, which generates massive numbers of redundant replicas that quickly exhaust the available buffer space across the network. Increasing the buffer size has minimal effect on reducing drops, because the replica growth rate increases faster than the available storage, illustrating the inherent inefficiency of the flooding-based approach. In contrast, Spray and Wait exhibits significantly lower drop rate between 1,000 and 5,700 messages, and benefits substantially from larger buffer sizes. Since the protocol limits the number of replicas, buffers do not become congested, allowing more messages to survive long enough to reach their destinations. The consistent reduction in dropped messages as the buffer increases further highlights the protocol’s robustness and scalability. 
\begin{center}
\includegraphics[width=0.8\columnwidth]{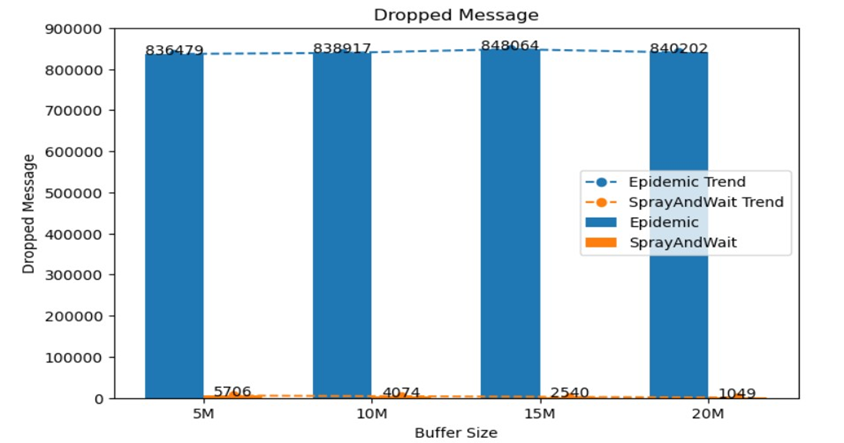}
\captionof{figure}{Dropped Messages in simulated football match scenarios.}
\label{fig1}
\end{center}

\section{Conclusion and Future Work}
\subsection{Conclusion}
This study explores a practical application of Delay-Tolerant Networks (DTN) in the case of emergency communications in a big football stadium under an earthquake incident. When conventional routes for mobile devices' data transmission, i.e., cellular infrastructure, are either unavailable due to physical distortion or blocked by traffic, DTN presents an alternative communication channel that enables mobile devices to transfer information opportunistically. To assess the efficacy of DTN-based routing, we designed a realistic simulation scenario within the ONE simulator, depicting six node types that were assigned different roles in the evacuation process: audiences, emergency response teams, ambulances, gathering points on the stage, reporters, and exit points. The two routing protocols, the Epidemic Protocol and the Spray and Wait Protocol, were evaluated depending on buffer size. 
 
The proof of concept has shown the clear distinction between the two protocols. Compared to the Spray and Wait method, the overall performance of the Epidemic protocol is characterized by lower delivery probabilities, higher latency, utilization of more hops, and elevated overheads to a large extent. Underlying this is the fact that excessive message duplication, buffer exhaustion, and acute packet loss rates, all of which make the Epidemic protocol most unsuitable for emergency scenarios, are also present. These results highlight the definition of routing strategies that retain the reliability of messages while optimizing the use of storage and bandwidth in environments that are suboptimal. The research thereby shows that methods of controlled replication routing, such as Spray and Wait, offer reliable communication solutions for crisis modes more in line with real conditions—those for which conventional network technologies are experiencing deficiencies.
\subsection{Future Work}
Several directions may extend the present study. Firstly, future work could broaden the routing comparison by including additional DTN protocols, such as PROPHET, MaxProp, or hybrid adaptive approaches like CafRep \cite{b15}, CafRepCache \cite{b10} and CognitiveCache \cite{b11} and so on, in order to determine whether the observed advantage of controlled replication remains consistent across a wider range of forwarding strategies. Secondly, energy efficiency should be considered more explicitly. In real disaster situations, mobile devices, rescue vehicles, and temporary communication nodes are usually battery-constrained. We plan to incorporate energy-aware opportunistic charging and energy distribution mechanisms, following the ideas proposed in sustainable vehicular edge and fog networks \cite{b16}. And as the next step, cross-layer communication design could be explored to improve coordination between routing, mobility, and network resource management in emergency environments \cite{b17}. Thirdly, we will incorporate our work in the Moditones platform and deploy it in future emergency real-world scenarios like school campus emergencies, partially connected communication environments, this would support the transition from simulator-based evaluation to low-cost practical prototyping \cite{b18}. Finally, more heterogeneous and realistic communication settings could be introduced, including drones, connected vehicles, and edge-assisted relay nodes for emergency earthquake scenarios, in order to reflect the dynamic communication structure often found in large-scale emergencies \cite{b19}.


\begin{thebibliography}{00}

\bibitem{b1} A.-C. Petre, C. Chilipirea, and C. Dobre, ``Delay tolerant networks for disaster scenarios,'' in \emph{Resource Management in Mobile Computing Environments}, 2014, pp. 3--24, doi: 10.1007/978-3-319-06704-9\_1.

\bibitem{b2} Wikipedia Contributors, ``August 2016 Central Italy earthquake,'' \emph{Wikipedia, The Free Encyclopedia}, Aug. 2025.

\bibitem{b3} Associated Press, ``Pescara vs. Atalanta halted as earthquake shakes stadium - ESPN,'' \emph{ESPN}, Oct. 2016. [Online]. Available: https://www.espn.co.uk/football/story/\_/id/37497007/pescara-vs-atalanta-halted-earthquake-shakes-stadium

\bibitem{b4} S. Kumar, ``A survey on delay tolerant network in disaster management,'' \emph{International Journal of Engineering Research \& Technology}, vol. 3, Jul. 2014, doi: 10.17577/IJERTV3IS070214.

\bibitem{b5} K. Fall, ``Retrospective on `A delay-tolerant network architecture for challenged internets','' \emph{ACM SIGCOMM Computer Communication Review}, vol. 49, pp. 75--76, Nov. 2019, doi: 10.1145/3371934.3371958.

\bibitem{b6} I. Chlamtac, M. Conti, and J. J.-N. Liu, ``Mobile ad hoc networking: imperatives and challenges,'' \emph{Ad Hoc Networks}, vol. 1, pp. 13--64, Jul. 2003, doi: 10.1016/S1570-8705(03)00013-1.

\bibitem{b7} A. Vahdat and D. Becker, ``Epidemic routing for partially-connected ad hoc networks,'' Tech. Rep.

\bibitem{b8} T. Spyropoulos, K. Psounis, and C. S. Raghavendra, ``Spray and wait,'' in \emph{Proc. 2005 ACM SIGCOMM Workshop on Delay-Tolerant Networking}, 2005, doi: 10.1145/1080139.1080143.

\bibitem{b9} A. Ker{\"a}nen, J. Ott, and T. K{\"a}rkk{\"a}inen, ``The ONE simulator for DTN protocol evaluation,'' in \emph{Proc. Second International ICST Conference on Simulation Tools and Techniques}, 2009, doi: 10.4108/ICST.SIMUTOOLS2009.5674.

\bibitem{b10} M. Radenkovic and A. Grundy, ``Efficient and adaptive congestion control for heterogeneous delay-tolerant networks,'' \emph{Ad Hoc Networks}, vol. 10, pp. 1322--1345, Sep. 2012, doi: 10.1016/j.adhoc.2012.03.013.

\bibitem{b11} M. Radenkovic and V. S. H. Huynh, ``Cognitive caching at the edges for mobile social community networks: A multi-agent deep reinforcement learning approach,'' \emph{IEEE Access}, vol. 8, pp. 179561--179574, 2020, doi: 10.1109/ACCESS.2020.3027707.

\bibitem{b12} J. T. B. Fajardo, K. Yasumoto, N. Shibata, W. Sun, and M. Ito, ``DTN-based data aggregation for timely information collection in disaster areas,'' in \emph{2012 IEEE 8th International Conference on Wireless and Mobile Computing, Networking and Communications}, pp. 333--340, Oct. 2012, doi: 10.1109/WiMOB.2012.6379095.

\bibitem{b13} N. Mehta and M. Shah, ``Performance of efficient routing protocol in delay tolerant network: A comparative survey,'' \emph{International Journal of Future Generation Communication and Networking}, vol. 7, pp. 151--158, Feb. 2014, doi: 10.14257/ijfgcn.2014.7.1.15.

\bibitem{b14} V. Kuppusamy, U. Thanthrige, A. Udugama, and A. F{\"o}rster, ``Evaluating forwarding protocols in opportunistic networks: Trends, advances, challenges and best practices,'' \emph{Future Internet}, vol. 11, p. 113, May 2019, doi: 10.3390/fi11050113.

\bibitem{b15} M. Radenkovic and A. Grundy, ``Framework for utility driven congestion control in delay tolerant opportunistic networks,'' in \emph{Proc. International Wireless Communications and Mobile Computing Conference}, pp. 448--454, Jul. 2011, doi: 10.1109/IWCMC.2011.5982575.

\bibitem{b16} M. Radenkovic and H. Huynh, ``Energy-aware opportunistic charging and energy distribution for sustainable vehicular edge and fog networks,'' in \emph{Proc. IEEE International Conference on Fog and Mobile Edge Computing (FMEC)}, Apr. 2020, doi: 10.1109/FMEC49853.2020.9144973.

\bibitem{b17} H. Huynh and M. Radenkovic, ``A novel cross-layer framework for large scale emergency communications,'' in \emph{Proc. International Wireless Communications and Mobile Computing Conference (IWCMC)}, Jun. 2017, doi: 10.1109/IWCMC.2017.7986616.

\bibitem{b18} M. Radenkovic, J. Crowcroft, and M. H. Rehmani, ``Towards low cost prototyping of mobile opportunistic disconnection tolerant networks and systems,'' \emph{IEEE Access}, vol. 4, pp. 5309--5321, 2016, doi: 10.1109/ACCESS.2016.2606501.

\bibitem{b19} M. Radenkovic, H. Huynh, R. John, and P. Manzoni, ``Enabling real-time communications and services in heterogeneous networks of drones and vehicles,'' in \emph{Proc. IEEE International Conference on Wireless and Mobile Computing, Networking and Communications (WiMob)}, Oct. 2019, doi: 10.1109/WiMOB.2019.8923246.

\end{thebibliography}
\end{document}